\documentstyle[12pt,epsf]{article}

\def\href#1#2{#2}  

\makeatletter
\@addtoreset{equation}{section}

\newcommand{\tr}{{\rm Tr}}
\newcommand{\beq}{\begin{equation}}
\newcommand{\eeq}{\end{equation}}

\newcommand{\I}{{\bf (I)}} 
\newcommand{\II}{{\bf (II)}}
\newcommand{\III}{{\bf (III)}}
\def\appendix{{\newpage\section*{Appendix}}\let\appendix\section%
        {\setcounter{section}{0}
        \gdef\thesection{\Alph{section}}}\section}

\begin{document}

\begin{titlepage}

\begin{flushright}
NSF-ITP-99-110\\
KUNS-1607\\
hep-th/9909202
\end{flushright}
%\vspace{12 mm}
\vfil\vfil\vfil\vfil

\begin{center}

{\Large {\bf Monopoles and Dyons in Non-Commutative Geometry}}

\vspace{10mm}

Akikazu Hashimoto$^a$ and Koji Hashimoto$^b$\\

\vspace{10mm}

$^a$Institute for Theoretical Physics\\ University of California,
Santa Barbara, CA 93106\\
aki@itp.ucsb.edu\\

\vspace{10mm}

$^b$Department of Physics\\
Kyoto University, Kyoto 606-8502, Japan\\
hasshan@gauge.scphys.kyoto-u.ac.jp\\

\end{center}

\vspace*{10mm}

\begin{abstract}
\noindent Taking advantage of the equivalence between supersymmetric
Yang-Mills theory on non-commutative spaces and the field theory limit
of D3-branes in the background of NSNS 2-form field, we investigate
the static properties of magnetic monopoles and dyons using brane
construction techniques. When parallel D3-branes are separated by
turning on a Higgs vacuum expectation value, D-strings will stretch
between them at an angle which depends on the value of the background
2-form potential.  These states preserve half of the supersymmetries
and have the same masses as their commutative counterparts in the
field theory limit. We also find stable $(p,q)$-dyons and string
junctions. 
%In the field theory limit, these preserve 1/2 and 1/4
%of the 16 supersymmetries respectively, and have appropriate
%corresponding masses. 
%We find that they do not preserve any supersymmetry but
%have the same masses as their commutative counterparts. In the field
%theory limit, the $(p,q)$-dyons and the string junctions restore 1/2
%and 1/4 of the 16 supersymmetries, respectively.
\end{abstract}

%\vspace{20mm}
\vfil
\begin{flushleft}
%\today
September 1999 
\end{flushleft}

\end{titlepage}

%%%%%%%%%%%%%%%%%%%%%%%%%%%%%%%%%%%%%%%%%%%%%%%%%%%%%%%%%%%%%%%%%%%%

\section{Introduction}

Quantum field theories on non-commutative geometries have received
renewed attention recently following the observation that they arise
naturally as a decoupled limit of open string dynamics on D-branes
\cite{Connes:1998cr}. In the formalism of \cite{Connes:1998cr},
supersymmetric Yang-Mills theory on non-commutative geometry (NCSYM)
arises from Fourier transforming the winding modes of D-branes living
in a transverse torus in the presence of NSNS 2-form background
\cite{Douglas:1998fm}. To be concrete, consider a D-string oriented
along the 01-plane and localized on a square torus in the 23-plane in
the background of $B_{23}$. In the absence of $B_{23}$, the Fourier
transform is equivalent to acting by T-duality in the
23-directions. In the presence of the $B_{23}$, however, the Fourier
transform \I\ and T-duality \II\ acts differently. On one hand, \I\
gives rise to the NCSYM with non-commutativity scale
\beq [x_\mu , x_\nu] = i \theta_{\mu \nu}. \eeq
On the other hand, \II\ gives rise to D3-branes in the NSNS 2-form
background. The precise map of degrees of freedom between \I\ and 
\II\ is highly non-local and was described in a recent paper
\cite{Seiberg:1999vs} as a perturbative series in the
non-commutativity parameter $\theta$. The physics of \I\ at large 't
Hooft coupling can further be related to \II\ in the near horizon
region \cite{Hashimoto:1999ut,Maldacena:1999mh} in the spirit of the
AdS/CFT correspondence \cite{Maldacena:1997re}.  Yet, these
equivalences have contributed very little to the understanding of the
localized observables in the NCSYM. The difficulty stems largely from
the fact that we do not yet understand the encoding of the observables
in one formulation in terms of the other with sufficient detail.

To study the localized structures, it is natural to introduce
localized probes. Topologically stable solution such as a magnetic
monopole seems particularly suited for such a task.  Instantons on
non-commutative space-times have also been studied
\cite{Nekrasov:1998ss} along this line.

In this article, we will study the static properties of magnetic
monopoles, dyons, and other related structures in the NCSYM with
${\cal N}=4$ supersymmetry\footnote{Related 1/2-BPS and 1/4-BPS
constant field-strength solutions on tori were discussed in 
\cite{Konechny:1998wv,Konechny:1999rz,Konechny:1999tu}.}. Since
the non-commutativity modifies the equation of motion for the gauge
fields, one must first establish the fact that these solutions exist
in the first place. To this end, the equivalence between \I\ and \II\
will prove to be extremely useful; magnetic monopoles and dyons can be
understood in \II\ in the language of brane configurations. Masses,
charges, and supersymmetries of these objects can be analyzed in the
language of \II. The fact that these objects stay in the spectrum of
the theory in the decoupling limit provides a strong evidence that
objects with the corresponding mass, charge, and supersymmetry exist
in the NCSYM. In the language of \II, it is also straightforward to
argue for the existence and stability of exotic dyons which arise from
three-string junctions
\cite{Aharony:1998bh,Dasgupta:1997pu,Sen:1998xi,Bergman:1997yw} and
other complicated brane configurations.

This paper is organized as follows. We will begin in section 2 by
briefly reviewing some basic facts about the NCSYM \I\ and how they
arise as a decoupling limit. Then we will take the magnetic monopole
as a concrete example and study its static properties in the language
of \II\ in section 3. In section 4, we will describe how the analysis
of section 3 can be generalized to $(p,q)$-dyons and string junctions.
We will conclude in section 5.

%%%%%%%%%%%%%%%%%%%%%%%%%%%%%%%%%%%%%%%%%%%%%%%%%%%%%%%%%%%%%%%%%%%%
%%%%%%%%%%%%%%%%%%%%%%%%%%%%%%%%%%%%%%%%%%%%%%%%%%%%%%%%%%%%%%%%%%%%

\section{Non-commutative Yang-Mills from String Theory}

In this section, we will review the string theory origin of the
NCSYM. To be specific, let us take our space-time to have 3+1
dimensions. We will not consider the effect of making time
non-commutative. Then, without loss of generality, we can restrict our
attention to the case where the only non-vanishing component of the
non-commutativity parameter is $\theta_{23} = -\theta_{32} = 2 \pi
\Delta^2$. ($\Delta$ has the dimension of length.) The NCSYM with
coupling $\hat{g}_{\rm YM}$ and non-commutativity $\theta_{\mu\nu}$ is
defined by the action
\beq S = \tr \int \! dx^4\, \left( {1 \over 4 \hat{g}_{\rm YM}^2} 
\hat{F}_{\mu\nu} * \hat{F}^{\mu \nu} + \ldots \right) 
\label{ncaction}\eeq
where ``$\ldots$'' corresponds to the scalar and the fermion terms,
$\hat{F}$ is the covariant field strength
\beq \hat{F}_{\mu \nu} = \partial_\mu \hat{A}_\nu - \partial_\nu
\hat{A}_\mu + \hat{A}_\mu * \hat{A}_\nu - \hat{A}_\nu * \hat{A}_\mu,
\eeq
and the $*$-product is defined by
\beq f(x) * g(x) = \left. e^{ i {\theta_{\mu \nu} \over 2}{\partial
\over \partial x_\mu} {\partial \over \partial x'_\nu} } f(x) g(x')
\right|_{x = x'}. \eeq
Relevant details about non-commutative geometry and the NCSYM are
reviewed in \cite{Connes:1998cr,Seiberg:1999vs}.

According to the construction of \cite{Connes:1998cr}, this theory is
equivalent to D3-branes in the background NSNS 2-form in the $\alpha'
\rightarrow 0$ limit while scaling
\beq g_{\rm s} = {1 \over 2 \pi} \hat{g}_{\rm YM}^2 
\sqrt{{\alpha'^2 \over
\alpha'^2+\Delta^4}},\qquad 
V_{23} = \Sigma_B^2 = {\alpha'^2 \over \alpha'^2+\Delta^4}\Sigma^2, \qquad 
B_{23} = {\Delta^2 \over \alpha'}, \label{scale}\eeq
and keeping $\Delta$, $\Sigma$ and $\hat{g}_{\rm YM}$ fixed. In the presence of
D-branes, longitudinally polarized constant NSNS 2-form is not a pure
gauge and has the effect of inducing a magnetic flux on the world
volume.  The magnetic fluxes in this context can be interpreted as the
non-threshold bound state of D-strings oriented along the 
1-direction. When multiple parallel D3-branes are present, the same
number of D-strings get induced on each of the D3-branes. When the
23-directions is compactified on a torus of size $\Sigma_B = \alpha'
\Sigma / \Delta^2$, the ratio of the number of induced D-strings and
the number of D3-branes is precisely $n_1 / n_3 = \Sigma^2 / 
\Delta^2$.

The map between gauge fields $\hat{A}_\mu$ of the NCSYM \I\ and the
gauge fields $A_\mu$ living on the D1-D3 bound state \II\ was
constructed in \cite{Seiberg:1999vs} to leading non-trivial order in
$\theta$, and takes the form
\beq \hat{A}_i = A_i - {1 \over 4} \theta^{kl} \{A_k, \partial_l A_i +
F_{li} \} + {\cal O} (\theta^2) \label{SWrel}\eeq
The resummation of this series is not well understood at the present
time\footnote{The higher order corrections to (\ref{SWrel}) were
studied recently in \cite{Asakawa:1999cu}.}.

%%%%%%%%%%%%%%%%%%%%%%%%%%%%%%%%%%%%%%%%%%%%%%%%%%%%%%%%%%%%%%%%%%%%
%%%%%%%%%%%%%%%%%%%%%%%%%%%%%%%%%%%%%%%%%%%%%%%%%%%%%%%%%%%%%%%%%%%%

\section{Magnetic Monopoles in NCSYM}

In this paper, we will study a variety of dyonic states in the NCSYM.
It will however be convenient to first study the case of the
BPS monopole as a prototype. The analysis for other cases will follow
a similar pattern.

%%%%%%%%%%%%%%%%%%%%%%%%%%%%%%%%%%%%%%%%%%%%%%%%%%%%%%%%%%%%%%%%%%%%

\subsection{Basic notions of the NCSYM monopoles}

We are interested in studying the properties of the monopole-like
objects in the NCSYM \I. To simplify our discussions, we will take our
gauge group to be $SU(2)$.  Some basic properties of the NCSYM action
is already manifest.  First, the $*$-product acts like an ordinary
product for the constant fields in the Cartan subalgebra of the gauge
group. Therefore, NCSYM can be Higgsed just like the ordinary
SYM. This is important since BPS monopoles exist as a stable state in
the Higgsed SYM.  Second, if we assume that only the magnetic field
and one component of the scalar (say $\hat \Phi_9$) is non-zero, the
terms in the action can be assembled into the form
\beq 
S = {1 \over 4 \hat{g}_{\rm YM}^2} {\rm Tr} \int\! dx^4  
\left[
\epsilon^{ijk}  
\left(  
\hat{F}_{ij} *D_k \hat{\Phi} 
+  D_k \hat{\Phi}  * \hat{F}_{ij} \right)
+  (\hat{F}_{ij} - {\epsilon_{ij}}^k  D_k \hat{\Phi} ) 
* (\hat{F}\rule{0ex}{1ex}^{ij} - \epsilon^{ijk}  D_k \hat{\Phi} ) 
\right].
\eeq
The second term in the integral is positive definite, so the action is
bounded below by
\beq S \ge 
 {1 \over 4 \hat{g}_{\rm YM}^2} {\rm Tr} \int dx^4\, 
\epsilon^{ijk}  \left(  
\hat{F}_{ij} *D_k \hat{\Phi} 
+  D_k \hat{\Phi}  * \hat{F}_{ij} \right) = 
 {1 \over 2 \hat{g}_{\rm YM}^2} {\rm Tr}  \int dx^4\,  
\partial_k \epsilon^{ijk}  \left(  
\hat{F}_{ij} * \hat{\Phi}  \right).
\eeq
Thus the notion of the BPS bound exists also in the non-commutative
theory. 
 
Now, by definition, a magnetic monopole solution should have the
property that
\beq \hat \Phi \rightarrow {U \over 2}\sigma^3 \eeq
at large $r$, so the bound on the action can be made to take the form
\beq S = {U \over 4 \hat{g}_{\rm YM}^2} {\rm Tr} \int_{S_2} dS_k\,
\epsilon^{ijk} \hat{F}_{ij} \sigma^3 \label{action2} . \eeq
Furthermore, in order for the action to be finite, $F_{ij}$ should
decay according to
\beq \hat B^k = \frac12\epsilon^{ijk} \hat{F}_{ij} = 
{x^k \sigma^3 \over 2r^3} Q \eeq
at sufficiently large $r$ where the system looks spherically
symmetric. Therefore, (\ref{action2}) is evaluated as
\beq S = {2 \pi Q \over \hat{g}_{\rm YM}^2} U .\eeq
In commutative theories, $Q$ takes on integer values due to the
Dirac's quantization condition. It is an important question whether
there are corrections to $Q$ in powers of $(\Delta U)$ for the
non-commutative theory.  Even in the non-commutative theory, however,
the fields are slowly varying for large enough $r$, so we expect the
standard commutative gauge invariance argument to hold. Therefore, we
are lead to conclude that the magnetic monopoles of NCSYM have the
same masses and charges as their commutative counterparts.

Here we have argued in general terms that a self-dual magnetic
monopole solution will saturate the BPS bound and has the same mass
and the charge as in the commutative theory, provided that they exist.
Unfortunately, the field equations of the non-commutative theory
contain an infinite series of higher derivative interactions, making
the task of proving the existence, as well as studying the detailed
structure of these solutions, a serious challenge. However, even
without the detailed understanding of magnetic monopole solutions in
NCSYM, the equivalence between \I\ and \II\ can be exploited to
establish some basic properties of these objects. For example, the
existence, the stability, the mass, and the supersymmetry of these
states can be understood in the language of brane construction in \II.
In this formalism, it is also easy to establish similar properties of
$(p,q)$-dyons and string junctions.  These brane constructions provide
a strong evidence that the corresponding objects exist in \I.

%%%%%%%%%%%%%%%%%%%%%%%%%%%%%%%%%%%%%%%%%%%%%%%%%%%%%%%%%%%%%%%%%%%%

\subsection{Brane construction of the NCSYM monopoles}

\begin{figure}[tdp]
\begin{center}
\parbox[b]{75mm}{
\begin{center}
\leavevmode
\epsfxsize=75mm
\put(-5,52){$x_9$}
\put(40,8){$x_1$}
\put(100,0){$\delta$}
\put(166,53){$2 \pi \alpha' U$}
\epsfbox{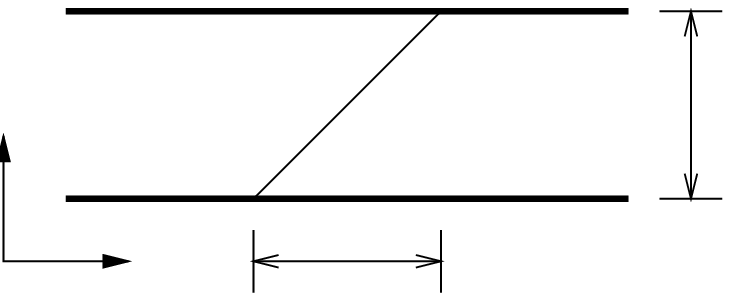}
\caption{Brane configuration of a D-string suspended between a pair of
parallel D3-branes in the background of the constant NSNS 2-form
\II. The induced magnetic field on the D3-brane world volume gives
rise to a tilt in the D-string orientation.}
\label{figa}
\end{center}}
\hspace{5mm}
\parbox[b]{75mm}{
\begin{center}
\leavevmode
\epsfxsize=35mm
\epsfbox{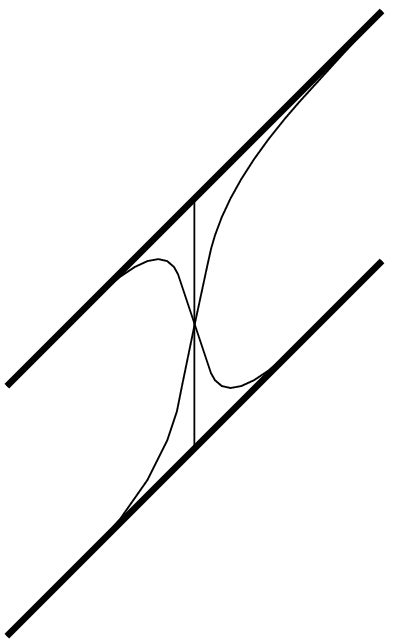}
\caption{Magnetic monopole solution in the tilted D3-brane picture
\III\ where the world volume fields are single-valued.}
\label{figb}
\end{center}}
\end{center}
\end{figure}

In the formalism of the field theory brane constructions, magnetic
monopoles in Higgsed SYM have a natural realization as D-strings
suspended between a pair of parallel but separated D3-branes. Similar
configuration exists in \II\ and is a natural candidate for a state
which gets mapped to the magnetic monopole of \I\ under the relation
(\ref{SWrel}).  One important difference between \II\ and the usual
situation is the fact that the background NSNS 2-form $B_{23}$ also
induces a background RR 2-form $A_{01} = {1 \over g} \sqrt{{B_{23}^2
\over 1+B_{23}^2}}$ which couples to the world volume of the suspended
D-string \cite{Maldacena:1999mh,Russo:1997if}. This effect can also be
interpreted as the force felt by the magnetic charge at the endpoint
of the suspended D-string in the background of constant magnetic field
in the 1-direction. The overall effect is to tilt the suspended
D-string in the 1-direction and to change the overall energy of the
configuration (see Figure \ref{figa}).
The extent of the tilt and the change in the energy can be found by
obtaining the minimal energy configuration of the D-string DBI
action in the RR 2-form background at weak string coupling
\beq S = {1 \over 2 \pi \alpha'} \int_0^{2 \pi \alpha' U} dx_9 
\left( {1 \over g_{\rm s}}
\sqrt{1 + \left({dx_1 \over dx_9}\right)^2} + A_{01}{dx_1 \over
dx_9}\right). \label{monopoleaction}\eeq
It is an elementary exercise to show that this expression is minimized
for $dx_1/dx_9 = B$, and that the minimum mass is 
\beq m = {U \over g_{\rm s}}\left(\sqrt{1+B^2} - {B^2 \over
\sqrt{1+B^2}}\right) = {2 \pi \over \hat{g}_{\rm YM}^2} U
\label{monopolemass}\eeq
where we used (\ref{scale}) to express the result in terms of the
parameters of the NCSYM \I. Despite the fact that the suspended
D-string was tilted in the 1-direction in response to the background
fields, the mass remained exactly the same as in the ordinary SYM.

It is also interesting to compute the ``non-locality'' of the
suspended D-string indicated by ``$\delta$'' in Figure \ref{figa}:
\beq \delta = {d x_1 \over d x_9} 2\pi \alpha' U = 2 \pi \Delta^2 U. \eeq
This length therefore remains constant in the decoupling limit
$\alpha' \rightarrow 0$ in spite of the fact that the slope
$dx_1/dx_9$ diverges in this limit.

It is straightforward to count the number of supersymmetries preserved
by this configuration. Let us denote the spinors representing 32
supercharges of type IIB theory by
\beq \epsilon_- = \epsilon_L - \epsilon_R, 
\qquad \epsilon_+ = \epsilon_L + \epsilon_R. 
\eeq
As we mentioned earlier, D3-branes in the background of $B_{23}$ can
be thought of as a bound state of $n_1$ D-strings and $n_3$
D3-branes. Such a configuration places a constraint
\beq \epsilon_- = \Gamma^0 \Gamma^1 
(\sin(\phi) \epsilon_- + \Gamma^2 \Gamma^3\cos(\phi) \epsilon_+) 
\label{d3susy} \eeq
on the supercharges, where $\tan(\phi) = B$. This result can be
easily obtained by following the supersymmetry of $(p,q)=(n_1,n_3)$
string through a chain of duality transformations.  On the other hand,
a D-string tilted in the 19-plane by the angle $\phi = \tan^{-1}(B)$
preserves
\beq \epsilon_- = \Gamma^0 \Gamma^\phi \epsilon_-, \qquad
\epsilon_+ = -\Gamma^0 \Gamma^\phi \epsilon_+ \label{d1susy}\eeq
where
\beq \Gamma^{\phi} = \Gamma^1 \sin(\phi) + \Gamma^9 \cos(\phi).
\eeq
The two constraints in (\ref{d1susy}) reduces the number of preserved
supersymmetries from 32 to 16. It turns out that (\ref{d3susy}) closes
among spinors satisfying (\ref{d1susy}), and reduce the number of
independent supersymmetries from 16 to 8. Therefore, this brane
configuration preserves the same number of supersymmetries as the
magnetic monopole of ${\cal N}=4$ SYM.

We are interested in the supersymmetry of these states in the field
theory limit where we scale $B = \Delta^2/\alpha' \rightarrow \infty$
keeping $\Delta$ fixed.  In this limit linear combinations of
(\ref{d3susy}) and (\ref{d1susy}) can be assembled  into the
following independent set of conditions
\begin{eqnarray}
&&\epsilon_- = \Gamma^0 \Gamma^1 \epsilon_-, \qquad 
\epsilon_+ = -\Gamma^0 \Gamma^1 \epsilon_+, 
\label{d3limitsusy}\\
&&\epsilon_- = \Gamma^9
\Gamma^1 \Gamma^2 \Gamma^3 \epsilon_+. \label{monopolesusy}
\end{eqnarray}
These conditions are satisfied by 8 spinor components, indicating
that the magnetic monopole preserves 8 out of 16 supercharges in the
field theory limit.

The brane configuration described in this section is precisely the
S-dual of the configuration considered in \cite{Hashimoto:1999xh},
except for the fact that in \cite{Hashimoto:1999xh}, it was the
D3-brane that was tilted instead of the D-string. The two description
can be mapped from one to the other by simply rotating the entire
system.  Although rotating the branes seem like a trivial operation,
it amounts to changing the static gauge condition in the language of
DBI action. The fact that this makes implicit reference to the
gravitational sector of the theory means that this is not a symmetry
in the field theory limit. It is more like a duality transformation
mapping equivalent physical system between two descriptions. Let us
therefore refer to the tilted D3-brane description as \III.

One particular advantage of \III\ is the fact that the field
configuration corresponding to this brane configuration is easily
understood.  Thinking of the pair of D3-branes as giving rise to $U(2)
= U(1) \times SU(2)$ gauge theory, the configuration of Figure
\ref{figb} is simply the $F_{23} = \partial \Phi_9 = B$ embedded into
the $U(1)$ sector and an ordinary Prasad-Sommerfield monopole embedded
into the $SU(2)$ sector \cite{Hashimoto:1998px}.

The equivalence between \II\ and \III\ also sheds light on the nature
of \II\ when expanded in $\theta$. When \III\ is interpreted as a
BIon, the fields are well defined as a single valued function. When
\III\ is rotated to \II, this single-valuedness is lost. The field
configuration must now contain branch cuts to account for
multi-valuedness in some region of the D3-brane world volume. Since
such a field configuration is non-analytic, expansion in $\theta$ is
likely not to yield a uniformly converging series, and this may have
profound implication for the map between \I\ and \II. Especially in
light of the fact that \II\ seem pathological from many points of
view, having a more conventional alternative description \III\ may
prove to be extremely useful in future investigations.

%%%%%%%%%%%%%%%%%%%%%%%%%%%%%%%%%%%%%%%%%%%%%%%%%%%%%%%%%%%%%%%%%%%%

\subsection{Magnetic monopoles at large N and large 't Hooft coupling}

Before concluding this section, let us pause for a moment and briefly
describe what happens to the magnetic monopoles in the NCSYM with
large 't Hooft coupling and large $N$. Consider $SU(N+1)$ broken to
$SU(N) \times U(1)$. At large coupling, this $SU(N)$ sector is
described by the supergravity background
\cite{Hashimoto:1999ut,Maldacena:1999mh} and the $U(1)$ sector appears
as a D3-brane probe in this background.  The supergravity background
describing the near horizon of the $N$ D3-branes in the background of
$B_{23}$ is given by
\begin{eqnarray}
ds^2 & = & \alpha' \left\{ \left( {U^2 \over \sqrt{\lambda}} \right) 
(-dt^2 + dx_1^2) + \left({\sqrt{\lambda} U^2 \over \lambda + 
\Delta^4 U^4}\right)(dx_2^2 + dx_3^2) + {\sqrt{\lambda} \over U^2}
dU^2 + \sqrt{\lambda} d \Omega^2 \right\}, \nonumber \\
e^{\phi} & = & {\hat{g}_{\rm YM}^2  \over 2 \pi}
\sqrt{{\lambda \over \lambda + \Delta^4 U^4}}, 
\qquad A_{01} = {2 \pi \over \hat{g}_{\rm YM}^2}
{ \alpha' \Delta^2 U^4 \over \lambda}, \qquad B_{23} = 
{ \alpha' \Delta^2 U^4 \over \lambda + \Delta^4 U^4},
\end{eqnarray}
where $\lambda = 4\pi \hat{g}_{\rm YM} N$. We wish to find the minimal
configuration for the probe D-string action
\beq S = 
{1 \over 2 \pi \alpha'} \int dx_1\left( e^{-\phi} \sqrt{-G_{00} (G_{11} +
G_{UU} (\partial U(x_1))^2)} - A_{01}\right) \label{d1action}.\eeq
Near the probe D3-brane, magnetic charge of the D-string will feel the
same force as in the case of the flat space, so we impose the boundary
condition that $\alpha' \partial U = \alpha' / \Delta^2$ at $U$ where
we place the probe D3-brane. Rather remarkably, the configuration
\beq U(x_1) = {1 \over \Delta^2} x_1, \eeq
{\it i.\,e.}\ a tilted straight line, is a solution to this problem,
and when the solution and the background is substituted into
(\ref{d1action}) we find
\beq S = \int dx {2 \pi \over \hat{g}_{\rm YM}^2} {1 \over \Delta^2} =
{2 \pi \over \hat{g}_{\rm YM}^2} U \eeq
which, as expected for a BPS state, is the same mass that we found in
the weakly coupled limit.

%%%%%%%%%%%%%%%%%%%%%%%%%%%%%%%%%%%%%%%%%%%%%%%%%%%%%%%%%%%%%%%%%%%%
%%%%%%%%%%%%%%%%%%%%%%%%%%%%%%%%%%%%%%%%%%%%%%%%%%%%%%%%%%%%%%%%%%%%

\section{{(p,q)}-Dyons and string junctions in NCSYM}

In the previous section, we described the interpretation of magnetic
monopoles of the NCSYM in the language of \II\ and found that they
have the same mass as the ordinary SYM. It is extremely
straightforward to repeat the analysis of the previous section to the
case of $(p,q)$-dyons. There will be some qualitative difference in
the pattern of supersymmetry breaking which we will discuss
below. Once the basic properties of the $(p,q)$-dyons are understood,
it is natural to consider the possibility of forming a state
corresponding to a string-junction
\cite{Aharony:1998bh,Dasgupta:1997pu,Sen:1998xi,Bergman:1997yw}. We
will examine the existence, the stability, and the supersymmetry of
these junction states.

%%%%%%%%%%%%%%%%%%%%%%%%%%%%%%%%%%%%%%%%%%%%%%%%%%%%%%%%%%%%%%%%%%%%

\subsection{(p,q)-Dyons in NCSYM}

It is extremely straightforward to generalize the discussion of the
previous section to the $(p,q)$-dyon. The expression for the action
(\ref{monopoleaction}) is generalized to
\beq S = {1 \over 2 \pi \alpha'} 
\int_0^{2 \pi \alpha' U} dx_9 \left( \sqrt{p^2 + {q^2 \over g_{\rm s}^2}}
\sqrt{1 + \left({dx_1 \over dx_9}\right)^2} +  q A_{01}{dx_1 \over
dx_9}\right). \label{dyonaction}\eeq
which is minimized by setting
\beq {dx_1 \over dx_9} = 
{q B \over \sqrt{(1+B^2)g_{\rm s}^2 p^2 + q^2}}.
\label{slope}\eeq
The minimum mass is 
\beq m = 
\sqrt{{(1+B^2) g_{\rm s}^2 p^2 + q^2 \over (1+B^2) g_{\rm s}^2}} 
U =  \sqrt{p^2 + {4 \pi^2 q^2 \over \hat{g}_{\rm YM}^2}} U
\label{dyonmass}\eeq
which is precisely identical to the result one would expect from the
ordinary SYM.

Let us now investigate the number of preserved supersymmetries for
these dyons. For the sake of concreteness, we will first consider
$(p,q) = (1,0)$, which is a W-boson.  As in the previous section, the
D3-brane puts the constraint (\ref{d3susy}). The $(1,0)$-string, on
the other hand, preserves
\beq \epsilon_- = \Gamma^0 \Gamma^9 \epsilon_+. \label{f1susy}\eeq
%
%The condition (\ref{f1susy}) will break half of the supersymmetries.
%
For spinors satisfying (\ref{f1susy}), the supersymmetry constraint
(\ref{d3susy}) simplifies to
%
%This time, spinors satisfying (\ref{f1susy}) do not automatically
%satisfy (\ref{d3susy}), but rather impose a new condition
%
\beq \left( 1 - \Gamma^0 \Gamma^1 \sin(\phi) - \Gamma^1 \Gamma^2
    \Gamma^3 \Gamma^9 \cos(\phi) \right) \epsilon_- =
    0. \label{susycond}\eeq
Conditions (\ref{f1susy}) and (\ref{susycond}) are satisfied by 8
independent spinor components for arbitrary values of
$\phi$.\footnote{We thank M. Krogh for pointing out an error regarding
this point in the earlier version of this paper.}  Therefore, we learn
that the W-boson in the field theory limit $\tan(\phi) = B =
\Delta^2/\alpha' \rightarrow \infty$ also preserves 8 supercharges.

%
%For general $\phi$, these two conditions leave 1/4 supersymmetries
%unbroken. 
%
%Generically, this condition does not have any solution, implying that 
%
%the $(1,0)$-string in the static equilibrium does not preserve any
%supersymmetry. 
%
%It is interesting to note, however, that 
%In the field theory limit, we
%are instructed to take $\tan(\phi) =B = \Delta^2 / \alpha'
%\rightarrow \infty$, and in that limit,  (\ref{susycond}) reduces
%to (\ref{d3limitsusy}), which is identical with the conditions for
%D-strings oriented along the 01-directions 
%(perpendicular to the fundamental string) and breaks half of the
%supersymmetries preserved by (\ref{f1susy}). Therefore the W-boson
%preserves 8 supercharges.

%On the other hand, supersymmetry constraints for the $(p,q)$-dyon with
%$q \neq 0$ are found to be similar to the ones for monopoles. This is
%due to the relation $p \ll q/g_{\rm s}$ in the decoupling
%limit. 
It is straight forward to extend this analysis to the case of $(p,q)$
dyons.  Taking the $(p,q)$-string to be oriented in the direction
given by (\ref{slope}), it is easy to obtain a set of independent
constraints in a manner similar to the monopole in the last
section. The number of the unbroken supersymmetries is 8, and in the
decoupling limit, the surviving supersymmetries are specified
%Before taking the decoupling limit, all supersymmetries are
%broken. However, in the limit, 8 supersymmetries survive, which are
%given 
by (\ref{d3limitsusy}) in addition to the constraint 
\begin{eqnarray}
\sqrt{p^2 + \left(\frac{2\pi q}{\hat{g}^2_{\rm YM}}\right)^2}\,
\Gamma^9 \epsilon_- =
\Gamma^1
\left(
  p + \frac{2\pi q}{\hat{g}^2_{\rm YM}}\Gamma^2 \Gamma^3
\right)\epsilon_+,
\label{pqsusy}
\end{eqnarray}
which reduces to (\ref{monopolesusy}) when 
$(p,q)=(0,1)$. 
We conclude that in the field theory limit, the $(p,q)$-dyons 
are 1/2 BPS objects, precisely analogous to the situation in the
ordinary ${\cal N}=4$ SYM.

Just as in the magnetic monopole case, one can consider the analogue
of \III\ where one tilts the D3-brane in such a way to make the
$(p,q)$-string point upward. This will simply correspond to embedding
the Julia-Zee dyon in the $SU(2)$ sector and turning on the $U(1)$
part independently. From this standpoint, it is easy to see that
the number of preserved supersymmetries is 8.

The large $N$ and large 't Hooft coupling limit of the $(p,q)$-dyon is
also straightforward to analyze. One simply generalizes
(\ref{d1action}) to
\beq S = {1 \over 2 \pi \alpha'} \int dx_1\left( \sqrt{p^2+q^2 e^{-2\phi}}
\sqrt{-G_{00} (G_{11} + G_{UU} (\partial U(x_1))^2)} - q A_{01}\right)
.\eeq
The minimal action configuration satisfying the appropriate boundary
condition is simply
\beq U(x) = {x \over \Delta^2} 
\sqrt{1 + {\hat{g}_{\rm YM}^4 p^2 \over 4 \pi^2 q^2}}, \eeq
and we find the mass of the $(p,q)$-dyon to be
\beq m =  U \sqrt{p^2 + {4 \pi^2 q^2 \over \hat{g}_{\rm YM}^4}}, \eeq
in agreement with the earlier result from weak coupling
(\ref{dyonmass}).

%%%%%%%%%%%%%%%%%%%%%%%%%%%%%%%%%%%%%%%%%%%%%%%%%%%%%%%%%%%%%%%%%%%%

\subsection{String junctions in NCSYM}

Having established the existence and some basic properties of
$(p,q)$-dyons, it is natural to consider the status of string
junctions.  In the absence of the background NSNS 2-form, the
existence of string junction relied on the property of
$(p,q)$-strings, that their tension can be balanced
\beq \sum_i \vec{T}_{p_i,q_i} = 0 \eeq
where 
\beq \vec{T}_{p,q} = \left( p ,{q \over g_{\rm s}}\right) \eeq
for $\sum p_i = \sum q_i = 0$. The components of $\vec{T}$ can be,
say, in the 8 and the 9 directions.

When the effect of the $B$-field is taken in to account, these vectors
are rotated out of the 89-plane into the 1-direction. Now one needs to
make sure that the tension balance condition is satisfied in the 1, 8,
and 9 directions simultaneously. It turns out, however, that the
entire effect of the $B$-field can be accounted for by rotating the
tension vector in the 19-plane so that the (1,8,9) components read
\beq \vec{T}_{p,q} = \left({q \over g_{\rm s}} 
\sin(\phi),\ p,\ {q \over g_{\rm s}}
\cos(\phi) \right), \qquad \tan(\phi) = B. \eeq
It is straightforward to verify that this vector is oriented relative
to the D3-brane world volume with the appropriate slope (\ref{slope})
by rotating $T_{p,q}$ in the 89-plane to point in the 19-directions.

Since we can just as easily tilt the D3-branes instead of tilting the
$(p,q)$-strings, there is a version of \III\ for the string junction.
The fact that the field configuration for such a state is known
\cite{Hashimoto:1998zs,Hashimoto:1998nj,Kawano:1998bp,Lee:1998nv}
might prove useful in the same way that the Prasad-Sommerfield
solution in \III\ is related to the magnetic monopole in the NCSYM \I.

Clearly, the condition for sum of $\vec{T}_{p,q}$ to vanish for
conserved $(p,q)$-charges in a string junction is still valid, so the 
string junction exists as a stable state in the presence of the $B$
field. Though diferent supersymmetries are preserved by the respective 
component $(p,q)$-strings in the string network, in view of this
stability the whole configuration is expected to preserve some of the 
supersymmetries.
%The supersymmetry of $(p,q)$-dyons are broken when these
%configurations are considered in the context of string theory in \II, 
%and the string junction must also break all supersymmetry.  However,
%the $(p,q)$-dyons restored their supersymmetry while keeping their
%masses finite in the field theory limit sending $\alpha' \rightarrow
%0$, and it would natural to expect similar restoration to take place
%in the string junctions. 
Let us therefore investigate the field theory
limit of these configurations more closely.

\begin{figure}[tdp]
\begin{center}
\leavevmode
\epsfxsize=50mm
\put(45,0){$x_9$}
\put(-3,45){$x_8$}
\put(50,35){$(-1,-1)$}
\put(50,120){$(1,0)$}
\put(100,63){$(0,1)$}
\epsfbox{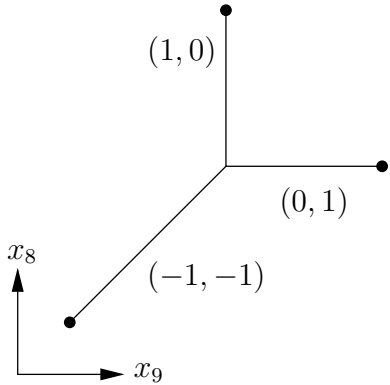}
\caption{Configuration of three string junction in a NSNS 2-form
background. The dots denote the D3-branes perpendicular to the
89-plane. The orientation of the branes resembles the conventional
junction in the 89-plane. The components of the junction is tilted in
the 19-plane in response to the NSNS 2-form background.}
\label{figc}
\end{center}
\end{figure}

Consider a junction of strings $(p_i, q_i)$, $i=1,2,3$, supported
by D3-branes localized in the 89-plane with strings meeting at the
origin. In order to take the field theory limit of such a
configuration, we should scale the distance of the D3-brane to the
origin as $\alpha' U_i$ with $\alpha' \rightarrow 0$  and
oriented in the $(p, {q \over g_{\rm s} \sqrt{1+B^2}})$ direction in
the 89-plane. In other words, the Higgs expectation value of the
$(\Phi_8,\Phi_9)$ field should be chosen to scale according to
\beq \vec{U}_i = (\Phi_8,\Phi_9)_i = 
{U_i \over \sqrt{p_i^2 + {q_i^2 \over (1+B^2)
 g_{\rm s}^2}}} \left(p_i,{q_i \over g_{\rm s} \sqrt{(1+B^2)}}\right).
\label{junchiggs} \eeq
To take the field theory limit, we scale $g_{\rm s}$ and $B$ according
to (\ref{scale}). Expressed in terms of $\hat{g}_{\rm YM}$ and
$\Delta$, (\ref{junchiggs}) reads
\beq \vec{U}_i = (\Phi_8,\Phi_9)_i = 
{U_i \over \sqrt{p_i^2 + {4 \pi^2
\over \hat{g}_{\rm YM}^4} q_i^2}} 
\left(p_i,{ 2 \pi \over \hat{g}_{\rm YM}^2}
q_i\right) \label{pqorientation}\eeq
and has a trivial $\alpha'\rightarrow 0$ limit. These junction states
therefore appear to exist in the field theory limit and orient itself
in the usual way in the 89-plane as we illustrate in Figure
\ref{figc}.
Figure \ref{figc} does not represent the orientation of the strings
outside the 89-plane but it should be remembered that they are tilted
in the 19-plane. The mass of the junction takes the same form as in
the commutative case
\beq
m=\sum_{i=1,2,3}
\sqrt{p_i^2 + {4 \pi^2 \over \hat{g}_{\rm YM}^4}q_i^2}
|\vec{U}_{p_i,q_i}|.
\eeq

The unbroken supersymmetries of the junction in the field theory limit
corresponds to the spinor components of the supercharges satisfying
the constraints of both the monopoles and the W-bosons,
(\ref{d3limitsusy}), (\ref{monopolesusy}), and (\ref{f1susy}). This
can be seen easily from the fact that, since the $(p_i,q_i)$-string is
now oriented in the direction (\ref{pqorientation}) in the 89-plane in
the decoupling limit, the constraint for the component
$(p_i,q_i)$-string becomes
\begin{eqnarray}
\left(p_i\Gamma^8 + \frac{2\pi q_i}{\hat{g}^2_{\rm YM}}\Gamma^9
\right) \epsilon_- =
\Gamma^1
\left(
  p_i + \frac{2\pi q_i}{\hat{g}^2_{\rm YM}}\Gamma^2 \Gamma^3
\right)\epsilon_+,
\end{eqnarray}
as a generalization of (\ref{pqsusy}).
We conclude, therefore, that objects in the NCSYM corresponding to the
field theory limit of the string junctions preserves 4 supercharges,
just like their commutative counterparts.

%%%%%%%%%%%%%%%%%%%%%%%%%%%%%%%%%%%%%%%%%%%%%%%%%%%%%%%%%%%%%%%%%%%%%%
%%%%%%%%%%%%%%%%%%%%%%%%%%%%%%%%%%%%%%%%%%%%%%%%%%%%%%%%%%%%%%%%%%%%%%

\section{Conclusions}

The goal of this paper was to understand the static properties of the
magnetic monopole solution and its cousins in the NCSYM. Instead of
working with the Lagrangian formulation of NCSYM \I, we took advantage
of the equivalence between NCSYM \I\ and the decoupling limit of
D3-branes in a background NSNS 2-form potential \II\ to study the
stable brane configurations corresponding to these states. Using this
approach, it is extremely easy to show that there are stable brane
configurations corresponding to magnetic monopoles, $(p,q)$-dyons, and
string junctions, and that they survive in the field theory limit. 

Having established some basic properties of these objects in the
language of brane construction, it is natural to wonder how much of
this can be understood strictly in the frame work of the Lagrangian
formalism.  It would be especially interesting to find an explicit
solution which generalizes the standard Prasad-Sommerfield solution
\cite{Prasad:1975kr} to the non-commutative setup. It is very
encouraging that the construction of instanton solutions via the ADHM
method admits a natural non-commutative generalization
\cite{Nekrasov:1998ss}. Indeed, Nahm's construction of the magnetic
monopole \cite{Nahm:1980yw,Nahm:1982jb,Corrigan:1984sv} also admits a
simple non-commutative generalization.  One simply solves for the
normalized zero modes of the operator
\beq 0={\bf \hat \Delta} ^{\dag}  * {\bf \hat v} = i {d \over dz} {\bf
\hat v}(z,x) - {\bf x} * {\bf \hat v}(z,x) - {\bf T} {\bf \hat v}(z,x)
\label{zeromode}, \eeq
and computes
\beq 
\hat{A}_i = \int_{-U/2}^{U/2} dz\, {\bf v}^\dag (z,x)  * \partial_i
{\bf v}(z,x), \qquad \hat{\Phi} = \int_{-U/2}^{U/2} dz\, z {\bf
v}^\dag (z,x) * {\bf v}(z,x).  \label{thesol}\eeq
The non-commutativity is reflected in the $*$-product in
(\ref{zeromode}), and as long as ${\bf \Delta}^{\dag} * {\bf \Delta}$
satisfies the usual requirement that it be invertible and that it
commutes with the quarternions, all the steps in the argument leading
to the self-duality of (\ref{thesol}) follow immediately from the same
argument in the commutative case \cite{Corrigan:1978ce,Nahm:1983sv}.
Despite tantalizing similarities with the commutative case, we were
not able to solve (\ref{zeromode}) in closed form to proceed
further. It would be very interesting to see if an explicit expression
for the non-commutative BPS monopole can be found.

%%%%%%%%%%%%%%%%%%%%%%%%%%%%%%%%%%%%%%%%%%%%%%%%%%%%%%%%%%%%%%%%%%%%%%
%%%%%%%%%%%%%%%%%%%%%%%%%%%%%%%%%%%%%%%%%%%%%%%%%%%%%%%%%%%%%%%%%%%%%%

\section*{Acknowledgments}

We would like to thank T.\ Asakawa, N.\ Itzhaki, I.\ Kishimoto, M.\
Krogh, and S.\ Moriyama for illuminating discussions. A part of this
work was carried out at the Summer Institute '99, Japan, and we thank
its participants for providing a stimulating working environment. The
work of A.\ H.\ is supported in part by the National Science
Foundation under Grant No.\ PHY94-07194. K.\ H.\ is supported in part
by Grant-in-Aid for Scientific Research from Ministry of Education,
Science, Sports and Culture of Japan (\#3160).

\begingroup\raggedright\endgroup

\end{document}